# Direct-write electrochemical nanofabrication of ultrasmall graphene devices

X. Liu[1] and C. Durkan[1,2]

[1]: Nanoscience, Department of Engineering, 11 JJ Thomson Avenue, Cambridge CB3 0FF, UK

[2]:corresponding author, email cd229@cam.ac.uk

**Abstract**

*Graphene nanoribbons (GNRs) are promising channel materials for next-generation ultra-miniaturized devices due to their exceptional electrical and thermal properties which arise from their atomic thickness, as well as their ability to have a size-dependent bandgap (1–9). However, despite extensive efforts to reliably fabricate narrow GNR-based field-effect transistors (FETs) (10–12), their integration into conventional transistor technologies remains hindered by challenges such as high fabrication costs and complex processing requirements(13, 14). In this study, we present a direct-write, relatively low-cost and robust approach for fabricating sub-10 nm GNR-based FETs using electrochemical atomic force microscopy (AFM) lithography with an alternating current (AC) bias, obviating the need for electrodes. We also explain the underlying electrochemical process and provide a model which can be used to describe it. Leveraging the high-precision positioning capability of AFM, this method enables precise nanoscale graphene patterning with feature sizes below 10nm. Compared with conventional lithographic techniques, photo- and electron-beam lithography (PL & EBL, respectively) (2, 15–20), it offers higher resolution, lower defect density, contamination-free processing, and the capability for in situ nanoscale device modification and characterization. This work provides an efficient strategy for advancing GNR-based nanoelectronics.*

## Introduction

The continued scaling-down of transistors in all 3 dimensions is the default approach that has been taken since the 1970s to reduce the cost and power consumption of highly integrated chips such as microprocessors (*21*). Moore's law has continued to hold true now for more than 50 years. Nonetheless, both theoretical and experimental works indicate that silicon transistors are rapidly approaching scaling limits due to a host of issues around short-channel, carrier mobility, defect density and quantum effects (*22*) as well as broader challenges around the manufacturability of nanoscale structures using conventional techniques (*23*) so we are in the early stages of the "beyond Moore" period. As a result, it is prudent to explore alternative materials and device fabrication methods with further scaling-down potential. The International Roadmap for Devices and Systems (IRDS) some time ago identified several promising alternatives to silicon, including III–V compound semiconductors, graphene nanoribbons (GNRs), and layered transition metal dichalcogenides (TMDCs) as well as a range of 2D materials (*24*). Among these, GNRs—narrow strips of graphene—are particularly attractive for logic microprocessors with advantages such as high, ambipolar carrier mobility, superior current-carrying capacity, and high saturation velocity (*25*). In order to be able to explore this space, reliable nanofabrication tools are needed which can be deployed to create test structures before any further work on scale-up can be explored. This is where the technique presented here fits in – in the exploratory phase of device fabrication.

For the last two decades, Photolithography (PL) and electron-beam lithography (EBL) have routinely been used to fabricate ultra-scaled FET devices based on GNRs (*2*, *15–20*). These techniques have several drawbacks including (i) requiring the use of multiple process steps involving chemicals and resists, (ii) limited precision, (iii) introduction of contamination and defects, (iv) long processing times, (v) harsh operating environments with high thermal budgets, and (vi) high fabrication cost. Atomic force microscopy (AFM)-based nanolithography shows promise as an alternative fabrication technique (*13*, *26–32*). In recent work, AFM-based lithography has been demonstrated under alternating current (AC) voltage conditions to pattern 2D materials. Liu et al. used this method to thin and etch black phosphorus (BP) on a SiO2/Si substrate (*33*).Subsequently, H. Li et al. demonstrated that graphene and boron nitride (hBN) / graphene heterojunctions can be patterned without leaving any residual material on the surface (*34*).

Despite the successful application of AFM-based lithography for etching materials, there remain many unanswered questions about (i) the underlying process and (ii) the nature of the contact between the tip and the sample which we address in this article. Li et al. assumed that a water-filled bridge existed between the tip and the sample, even under contact-mode operation. Thus, they considered the AFM-based lithography patterning technique with AC bias to be the same as oxidation scanning probe lithography (o-SPL) under non-contact mode. We will show that even with the tip in contact with a graphene surface, large enough local electric fields can be generated to initiate the etching process. Another critical factor in determining the success of this approach is the relative humidity (RH) and the subsequent capillary condensation around the tip-sample junction.

In this study, we systematically investigate AFM-based lithography utilizing an AC bias to pattern GNRs. We examine the key parameters influencing the process, including the applied force on the tip, tip velocity, the amplitude and frequency of the voltage applied to the tip, and the relative humidity (RH). Our results demonstrate that the direct contact between the tip and sample surface, along with the lateral voltage drop across the adjacent graphene surrounding the tip, plays a critical role in this technique. Additionally, we have developed a finite-difference electrostatic model to explore the electric field and potential distribution in the vicinity of the tip-graphene contact, which takes into account the influence of the adsorbed water, which is the electrochemical feedstock. Additionally, we have modelled the adsorption and meniscus formation processes which creates the nanoscale reaction environment around the tip apex. We also show that the complex permittivity of the adsorbed water plays a key role in determining the frequency dependence of the efficacy of the technique, and why it does not work under DC conditions. Additionally, we show how the lateral size of the graphene plays a role in determining the efficacy of the technique, via its capacitance. We therefore refer to the technique as AC-LAO. This work paves the way for the fabrication of next-generation FETs based on two-dimensional materials, facilitating both fundamental studies and device prototyping.

## Results and Discussion

Figure 1 (a) shows a schematic illustrating the concept of AFM-based lithography using an AC bias. The AFM was operated in contact-mode, with a relative humidity (RH) level of over 45%. Under these conditions, a water meniscus forms spontaneously in the high-curvature region around the contact point between the tip and the sample due to capillary condensation (*35*, *36*). Our experiments show that etching of the graphene occurs within this meniscuswhen the applied voltage exceeds a threshold value of approximately 5.5V. For RH levels of below 35%, we have been unable to demonstrate any etching even with an amplitude of the applied voltage up to 10V. Figure 1(b) shows a schematic illustrating the nature of the environment around the tip-sample junction. The graphene is not electrically connected to anything, so is essentially floating. It is worth noting the following points:

- No etching occurs under DC conditions for either a positive or negative tip voltage
- Etching works reliably under the application of an AC voltage when the graphene is electrically floating
- When the graphene is grounded, no etching takes place

Figure 1 (c) shows three etching trenches that we fabricated on a CVD graphene sample in this way. The frequency and amplitude of AC voltage were set at 20 kHz and 10 volts (peak voltage) separately. The contact force was set at 60 nN.

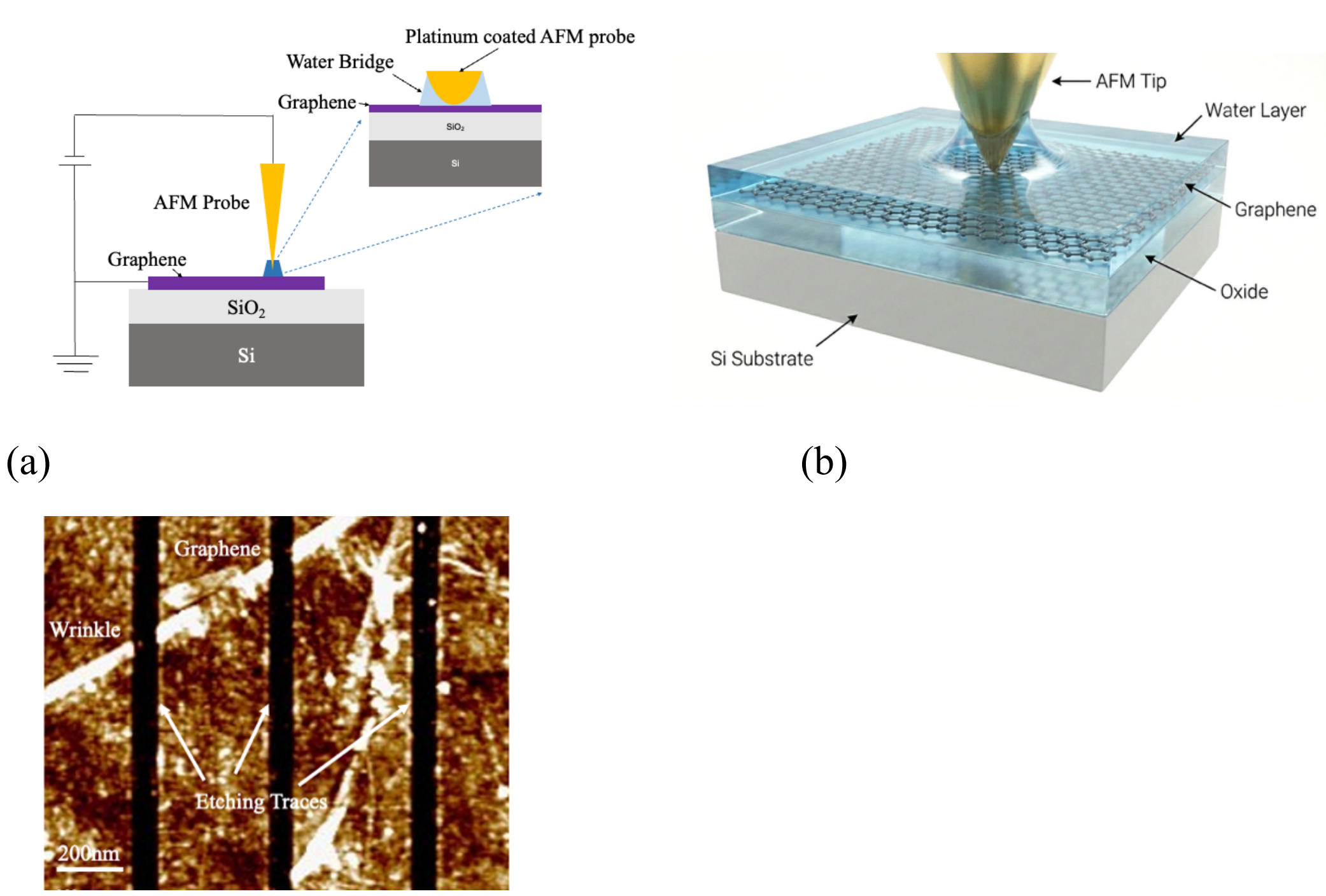

(c)

*Figure 1*. *(a) Schematic of AC-Local Anodic Oxidation (LAO), whereby an AC voltage is applied to a conductive AFM tip, which is in contact with an electrically isolated graphene layer in a high humidity environment; with a zoomed-in image of the contact; (b) illustration of the nature of the adsorbed water around the tip-sample contact (c) An array of etched lines fabricated by AC-LAO.*

Having experimentally determined that graphene can be reliably etched at a frequency of 20 kHz and an amplitude of 10V, the impact of the contact force was examined. The etching results with forces in the range 0nN-to 100nN are shown in Figure 2. Etching was not successful with a contact force of 0nN, as shown in Figure 2 (a), indicating that the tip and sample need to be in stable contact in order for this technique to be successful. Furthermore, although the size of the etched region is independent of the contact force in this range, it is clear that etching with larger forces has the additional effect of removing any debris within the etched region.

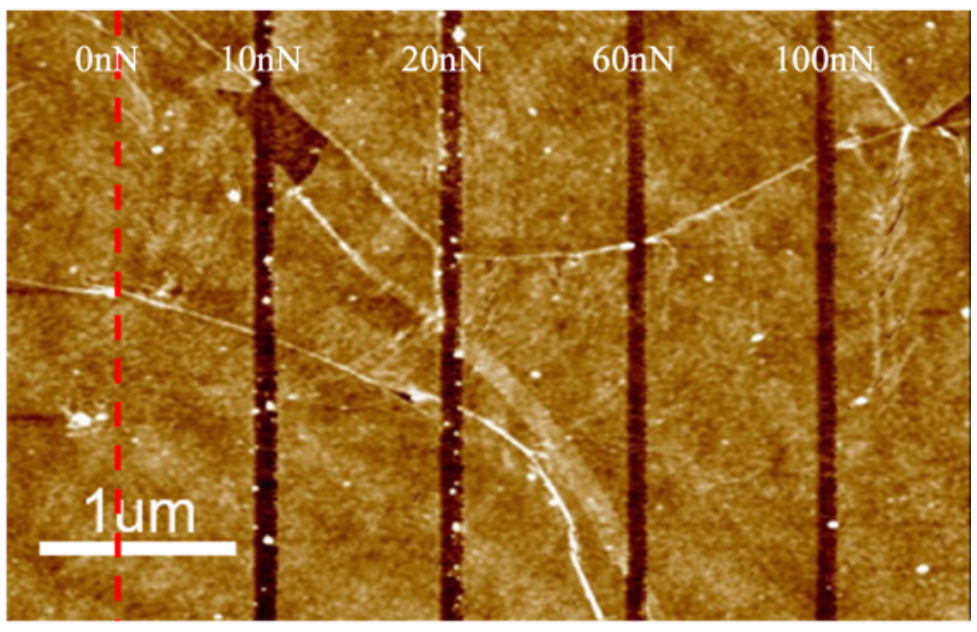

***Figure 2*** *The influence of the tip-sample contact force on the etching effect. Topography image of etched lines under different set points: 0nN, 10nN, 20nN, 60nN and 100nN*

The etching process occurs as a result of an electrochemical reaction enabled by the presence of water, mediated by the applied voltage, the corresponding electric field and the current flowing into the graphene. This reaction will therefore be highly voltage-dependent, and there will be a threshold voltage below which there is insufficient energy to break the C-C bonds within graphene. To determine the value of this threshold, a series of patterns were programmed with the tip AC voltage stepping up from 1V peak amplitude to 9V. The results are shown in Figure 3, where it is clear that this threshold is between 5V and 5.5V. When the applied voltage was increased beyond this, the linewidths of the etched regions also increased, as shown in Figure 3. We can also clearly see that the minimum linewidth is of order 23 nm.

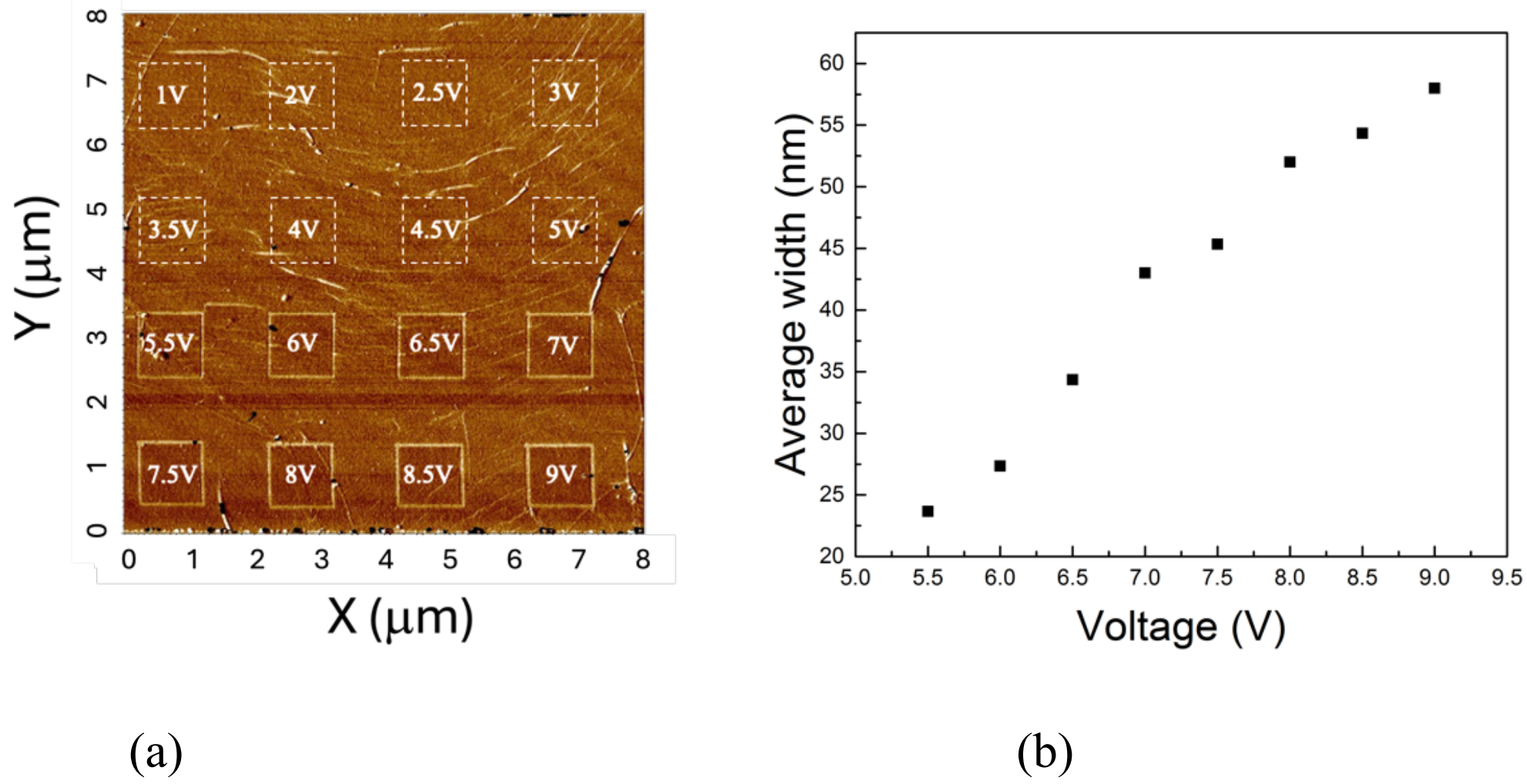

***Figure 3.*** *Variation of etched linewidth as a function of the amplitude of the applied AC voltage to the tip (at 20 kHz) for programmed square-frame patterns. (a) AFM Phase image showing that no patterning occurs below 5.5V. (b) Plot of the relationship between voltage amplitude and the linewidth of the etched regions.*

The adsorbed water will inevitably contain impurities, leading to it having a finite conductivity and therefore acting as an electrolyte. The ions present in the water form a double layer on the surface of the graphene, and the rate at which the graphene etches is determined by the current flowing into the it via this double layer, so one would expect that the dwell time of the tip at any location will influence the extent of the etching. To examine this, the effect of the lateral speed of the AFM tip as it moved across the surface during application of the AC voltage was examined. The range of tip speeds was varied from 1nm/s up to 1 μm/s. The AC voltage had a peak amplitude of 8V at a frequency of 20 kHz, meaning that within a 1nm area, which is the approximate size of the contact radius, the number of voltage cycles experienced by the sample immediately under the tip was in the range 20,000 to 20. Higher scan speeds led to incomplete patterns, partly as a result of the difficulty of maintaining good feedback control and also due to the reduced number of AC cycles which took place at the contact point.

In the range of tip speeds up to 2.5nm/s, as well as the graphene being removed over a wider area, up to around 115nm across, there is evidence of the deposition/growth of material in the newly-exposed area, as shown in Figure 4. This is as a result of the growth of an oxide layer via a LAO process on $SiO_2$ (*37*), which is more usually associated with the application of a negative DC voltage on the tip. The thickness of this deposit is seen to vary dramatically with tip speed in that it is clearly visible and over 3nm thick at a speed of 1nm/s, and appears to no longer be present by the time the speed is 2.5nm/s as shown in Figure 4.

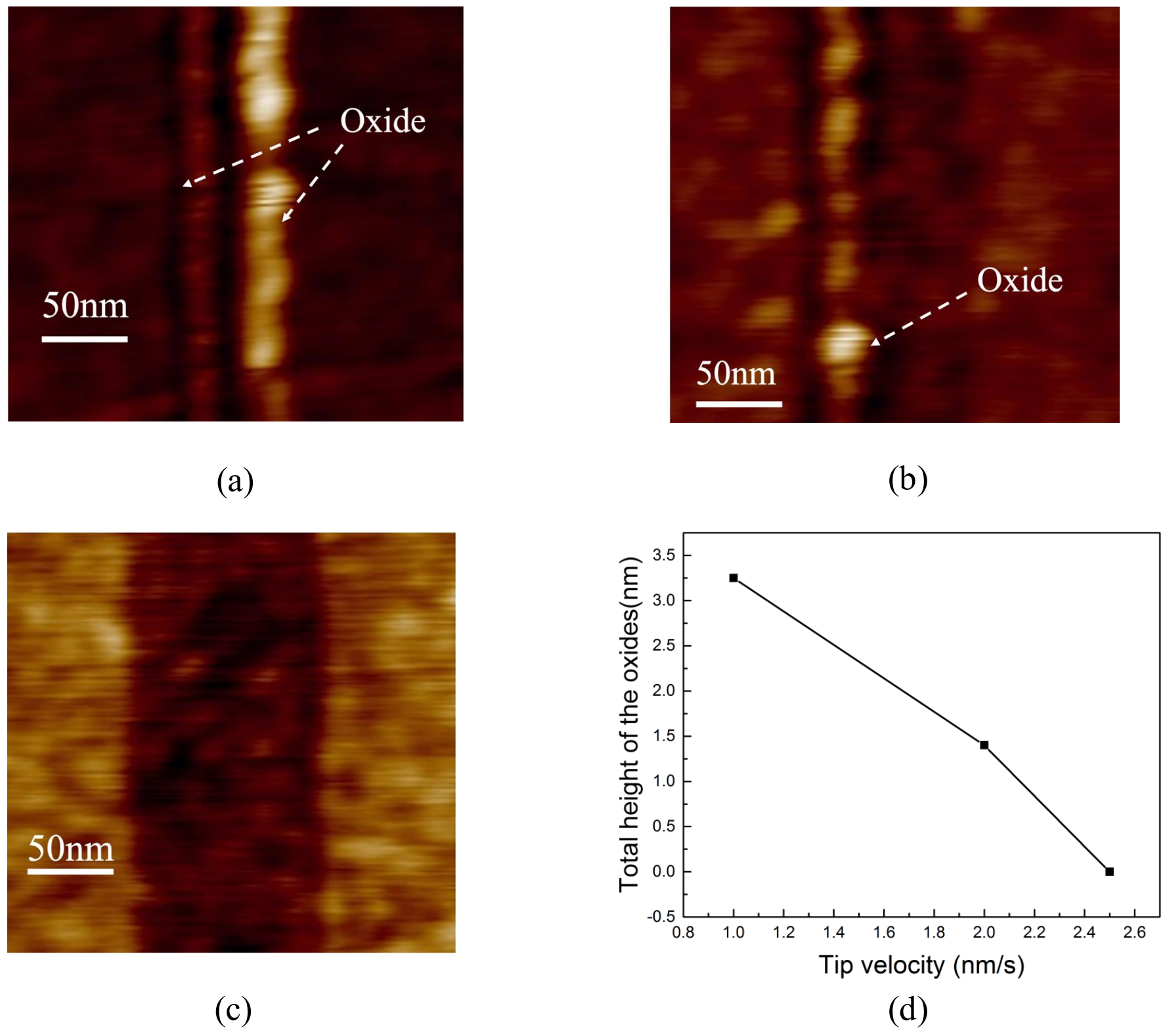

***Figure 4***. *for very low tip speeds, oxide protrusions are seen to grow on the surface where graphene has been removed. Topography images showing this growth for tip speeds of (a) 1nm/s (b) 2nm/s (c) 2.5nm/s (d) measured height of oxides as a function of tip velocity.*

In the higher-velocity regime above 2.5nm/s, where the dwell time is relatively shorter, the etched regions were clean without any noticeable oxides and their width was seen to decrease with increasing tip speed up to a speed of around 500 nm/s, as shown in Figure 5.

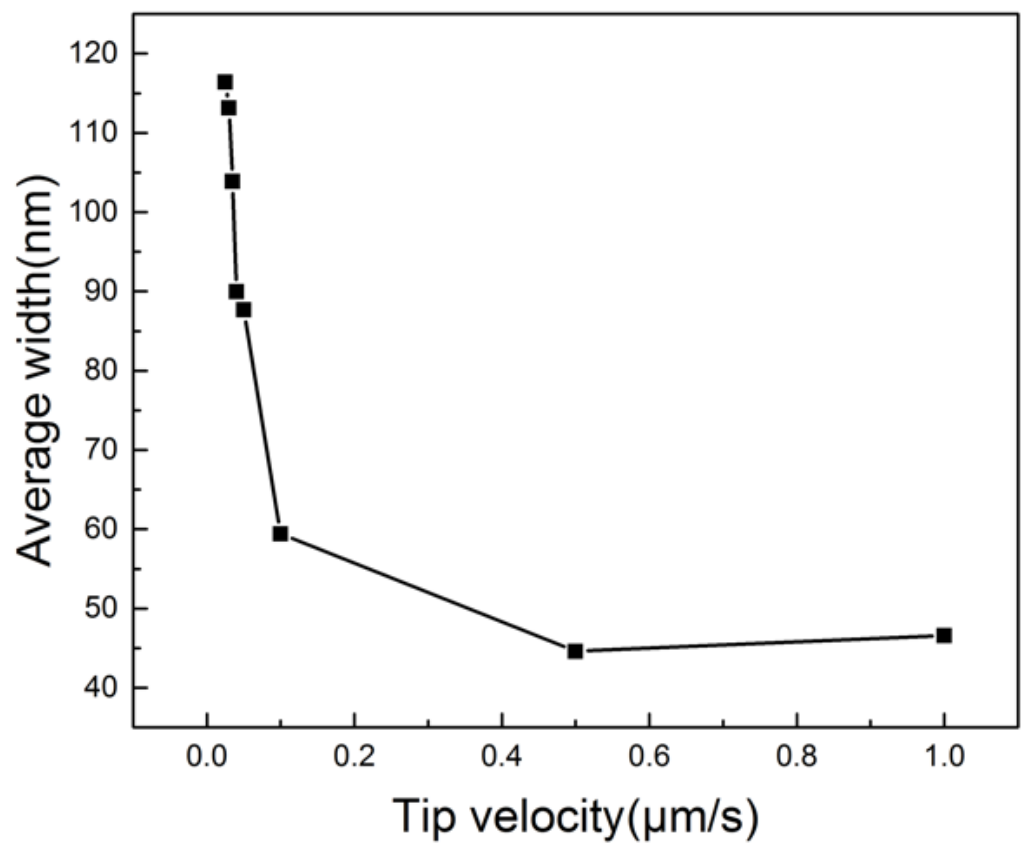

***Figure 5.*** *Illustration of the variation of the width of the etched lines as the tip velocity increases from 2.5nm/s to 1μm/s.*

In the range of frequencies which can be applied to the tip in this AFM system, which is 100 Hz to 600 kHz, the size of etched structures is independent of frequency, as long as the graphene being used is more than several tens of microns across. However, higher frequencies give rise to better-defined and cleaner edges.

Therefore, to achieve reliable etching with the highest resolution in order to enable the fabrication of nanoscale devices, the contact force, AC voltage amplitude and frequency and tip speed were thereafter chosen as 60 nN, 5.5 V, 600 kHz and 1 µm/s respectively. In Figure 6, we show a series of the narrowest structures we have been reliably able to create, which are approx. 24nm wide. Figure 6(a) shows the topography where the etched regions are clearly demarcated, and in Figure 6(b) which is a phase image, the phase is clearly positively shifted in the etched regions, which is what is usually observed with $SiO_2$. This phase shift, combined with the reduction in height in the exposed regions is clear evidence that there are two different materials, consistent with the removal of graphene to expose the underlying $SiO_2$. In Figures 6(c) and 6(d), we show some other structures that have been written to illustrate the achievable resolution.

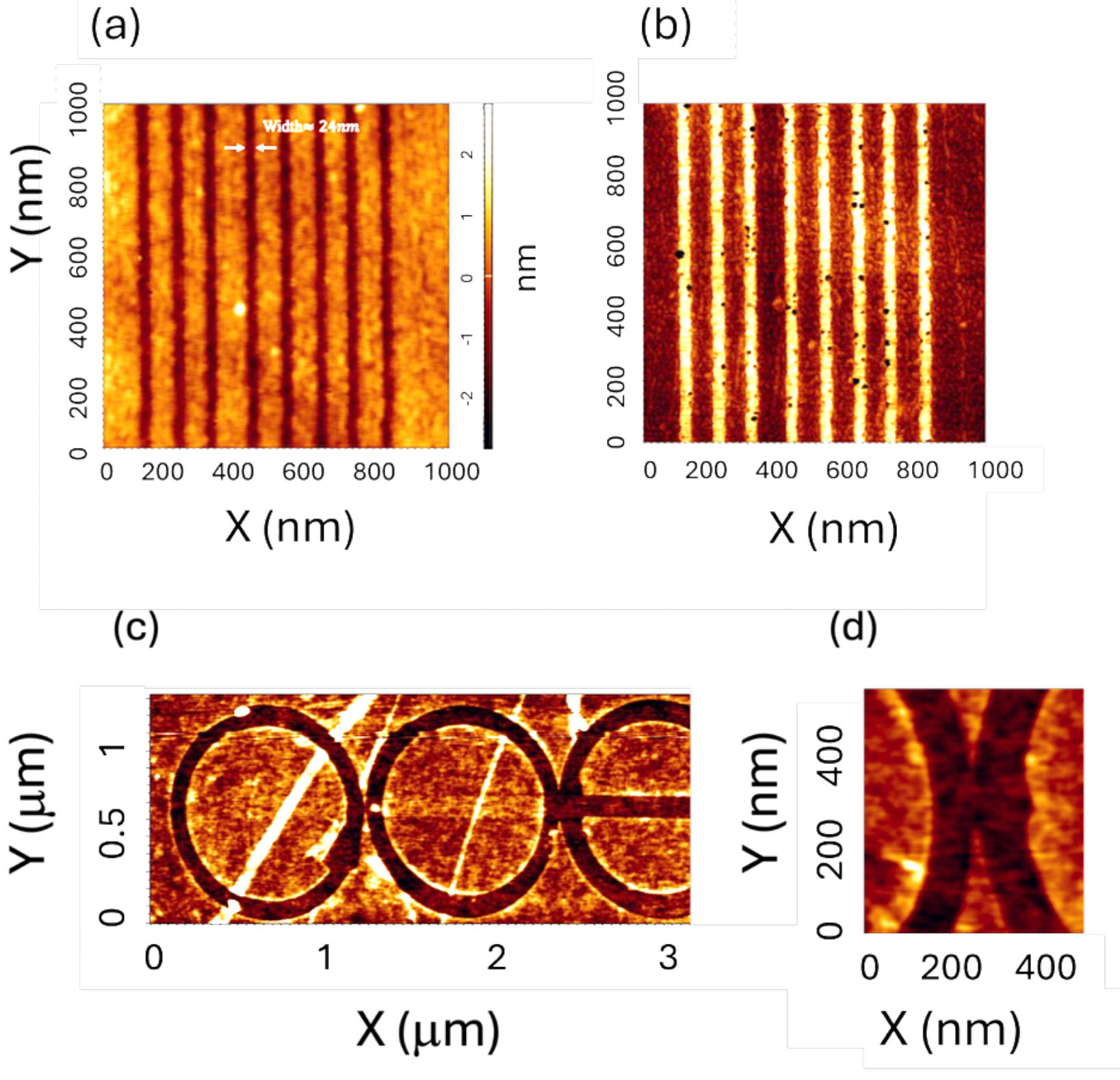

***Figure 6.*** *The smallest linewidths etched by AC-LAO. (a) topography image of the etching results with width which is around 24nm. (b) The phase image of the same areas. The bright areas show the exposed* $SiO_2$*; (c) A series of circles patterned close enough together to create nanogap structures, a zoom-in of one of which is shown in (d).*

In order to further confirm the removal of graphene and that this AFM-based lithography process is potentially an appropriate technique for the fabrication of nanoscale devices, Conducting AFM (C-AFM) was used to verify that the etched regions are indeed insulating. A series of square-frame patterns were programmed with 5 different tip voltages: 0V, 3V, 5V, 7V and 10V. The same region was then imaged twice: once in tapping mode where the topography and phase were collected and a second time in contact mode, where the sample was connected to a DC voltage of 100 mV and the resulting current was collected by a transimpedance amplifier connected to the tip (i.e. in C-AFM mode). The resulting phase and current map images are shown in Fig. 7 (a) and (b). At and above the threshold value of around 5V, the graphene was removed as we can (i) see the difference in the phase signal again indicating different materials, and (ii) from the current maps, the fabricated isolated graphene islands surrounded by the etched areas are clearly electrically isolated, so the etched regions are indeed insulating.

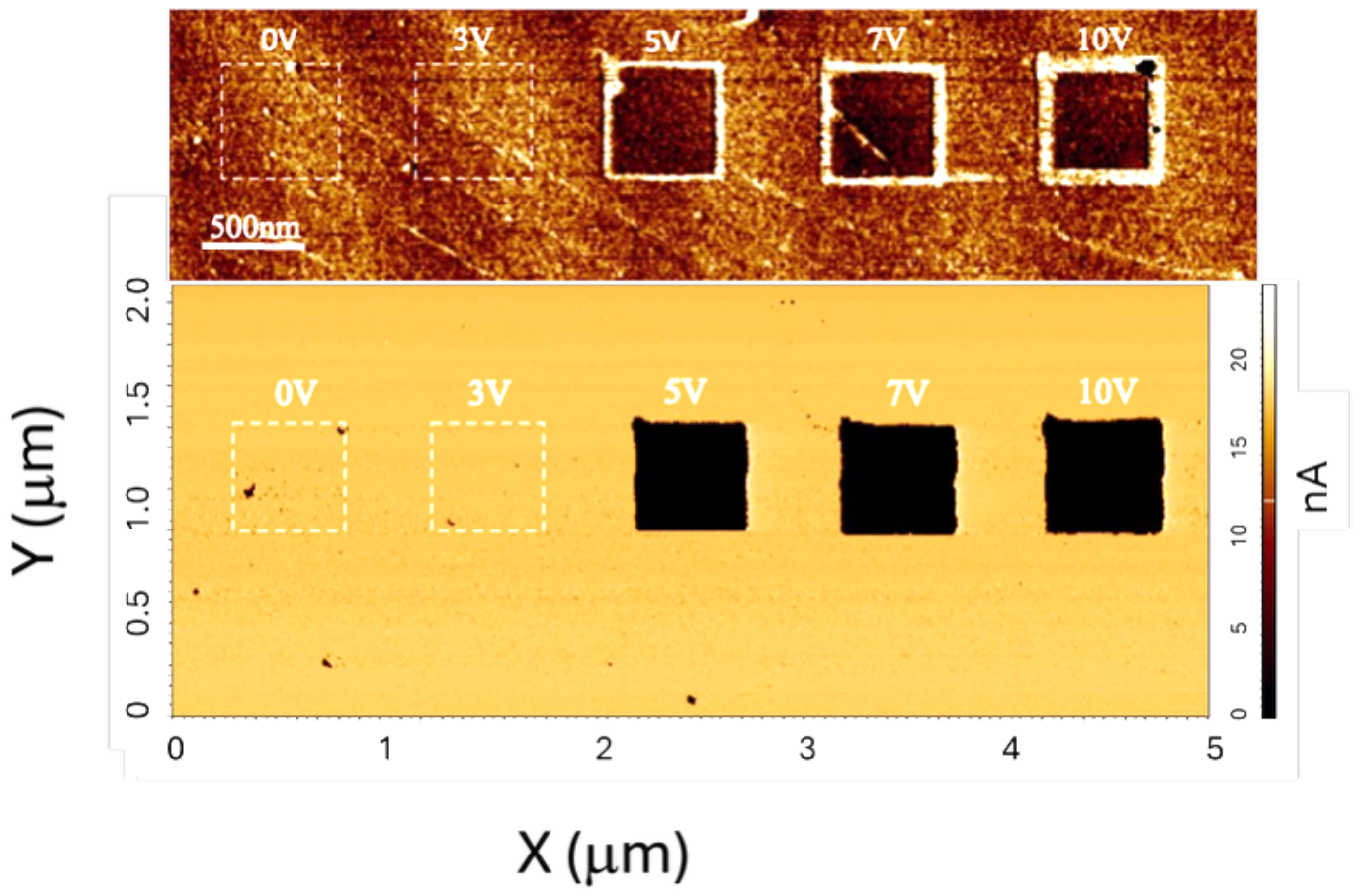

***Figure 7.*** *The phase image and current mapping results of o-SPL etching for a series of voltages. (a) The phase image scanned by AFM. This image was scanned by AFM in tapping mode. (b) current maps taken simultaneously with topography, in C-AFM mode.*

Given that the electric field around the tip will depend on the size of the graphene due to fringing effects, an isolated island of graphene approx. 1.7 microns across was first created, and an attempt was made to etch inside it. The result of this two-step etching experiment is shown in Figure 8. The contact force, frequency and amplitude of the voltage was set at 60nN, 20kHz and 10V. In step 1, a perimeter was etched, starting from and then returning to point A and following the solid arrow. For Step 2, the starting point, A' was located towards the middle of the isolated graphene area that had just been created, as indicated by the black solid arrow and dot. The tip followed the dotted arrow path shown, and as can be seen, the etching process failed to cut graphene from A' until the tip exited the isolated graphene fabricated by Step 1. It is important to note that this phenomenon cannot be explained by the mechanism of o-SPL proposed by X. Liu et al (*33*). In their explanation, the tip is not in contact with graphene, leaving a gap between them which is filled by water. In this way, there would be no difference between step 1 and 2 and the etching process should still work from A' to the boundary of Step 1.

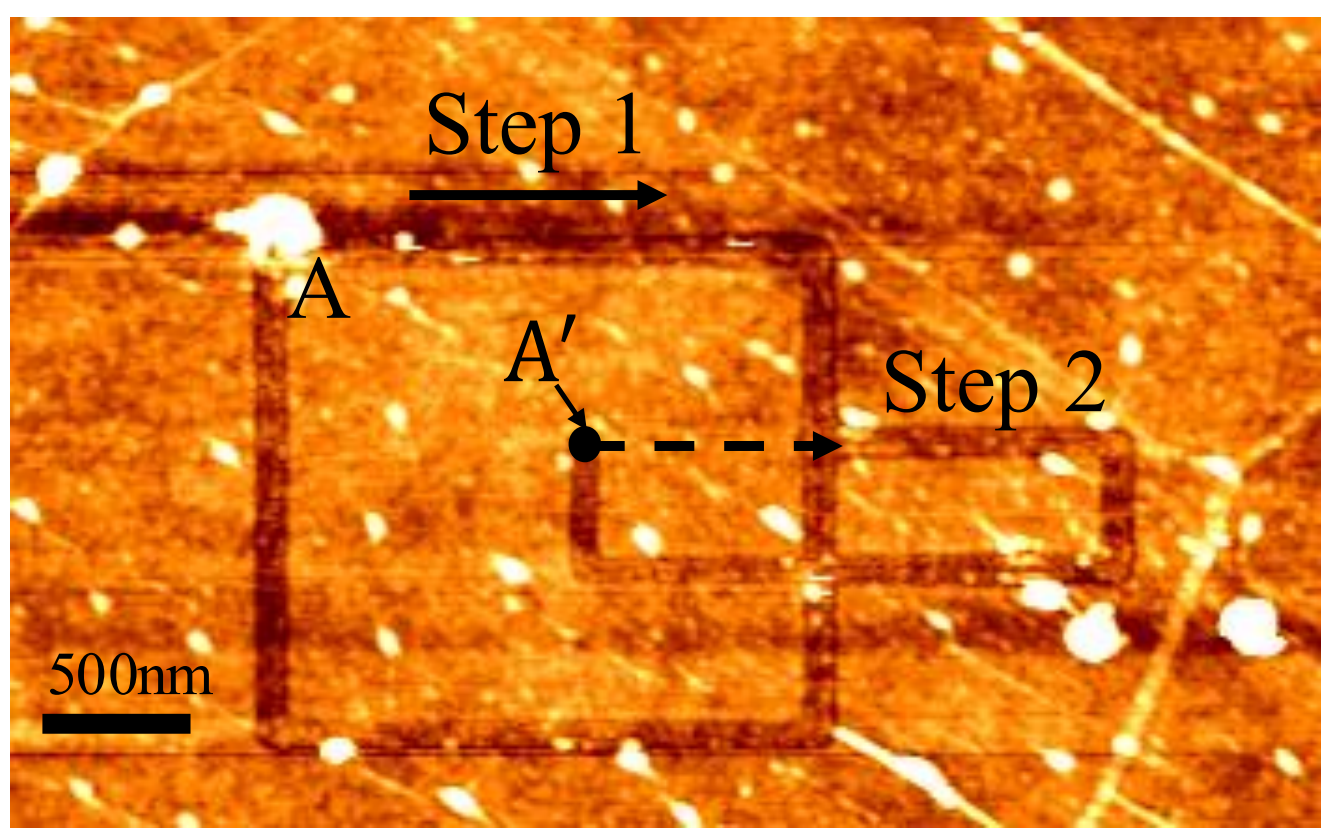

***Figure 8.*** *Topography image of a two-step etching. This image was scanned by AFM in non-contact mode In Step 1, etching was initiated at point A (indicated by the black arrow) to define a square pattern. In Step 2, etching was subsequently initiated at A′ to define a second square pattern. The dashed line indicates a failed etching path.*

The failure of the etching process as the tip moves from A′ to the boundary, followed by successful patterning beyond the boundary, indicates that the peak electric field near the tip changes when the boundary is crossed. In the next section, we will develop a model to understand the underlying processes which are occurring in the vicinity of the contact between the tip and the graphene, and which can explain our observations.

In order to gain an understanding of the underlying process and the influence of experimentally-controllable parameters, we propose an equivalent circuit diagram of the tip-graphene configuration as shown in Figure 9. The resistance of the tip itself is of order kOhms, so is negligible compared to the other resistances, but nonetheless should be included for completeness. We have already determined that the contact resistance between the tip and the graphene, $R_{contact}$ is of order at least 2 MΩ. This is the dominant pathway through which current will flow between the tip and graphene, but it has no bearing on the etching process. There is then a parallel pathway for current to flow between the sidewalls of the tip and the graphene, through the water in the meniscus, aided by the flow of the ions as discussed above. There will also be a few monolayers of water adsorbed across the entire graphene surface, but the sheet resistance of this will be high enough that the current expected to flow laterally through it will be negligible, i.e. the current will be largely confined within the meniscus.

The current localised within the meniscus is expressed as a parallel combination of a resistor, $R_{water}$ and capacitor, $C_{water}$. In series with this is the impedance associated with the double layer on the graphene surface within the meniscus, immediately above the graphene, of order 1nm or so thick, which has capacitance $C_{dl}$ (typically of order 20-40 μF/cm$^2$). Then, the graphene/oxide/back electrode system itself has capacitance and leakage resistance associated with the oxide layer, modelled as $C_{ox}$ and $R_{leakage}$.

We also note that the impedance of the graphene island (via $C_{ox}$) will scale inversely with area, so the etching rate will be lower for smaller islands for a given AC frequency. This is essentially what we are seeing in the experiment in Figure 8. We would also expect that the scaling with area is only approximate, and the dependence on size will start to deviate from a simple linear relationship once the lateral size of the graphene becomes comparable with the oxide thickness, due to the emergence of fringing effects.

Under DC conditions, we can see from the equivalent circuit therefore that the capacitance associated with the double layer essentially acts to block all current flowing through the water, apart from an initial spike as the double layer charges up. The only current flowing through the system is therefore just whatever passes through the contact point, and no etching will take place. Likewise if the graphene is grounded, where we experimentally observe the etching no longer works, the pathway via the graphene is short-circuited, which essentially pulls down the voltage appearing across the meniscus. In order to be able to estimate the voltage on the graphene and the frequency behaviour of the system, we need an estimate of each of the component values, as shown in Table 1.

| **Component** | **Value** | **Notes** |
|---|---|---|
| $R_{tip}$ | kΩ | This is the resistance of the current pathway along the tip and cantilever through the metal coating |
| $R_{contact}$ | 2MΩ | The contact resistance of the AFM tip with the graphene sample – experimentally determined to be of order 2 MΩ |
| $C_{water}$ | aF | |
| $R_{water}$ | ~ 1-10 GΩ | The Kelvin radius is of order 18nm, and the height of the meniscus is around 1 nm, yielding a resistance of 1 GΩ for a conductivity of $10^{-3}$ S/m |
| $C_{dl}$ | 20-40 μF/cm$^2$ | |
| $C_{ox}$ | Of order aF | Depends on graphene size |
| $R_{leakage}$ | Of order $10^{14}$Ω or more | Depends on graphene size |

**Table 1.** Parameters for the equivalent circuit.

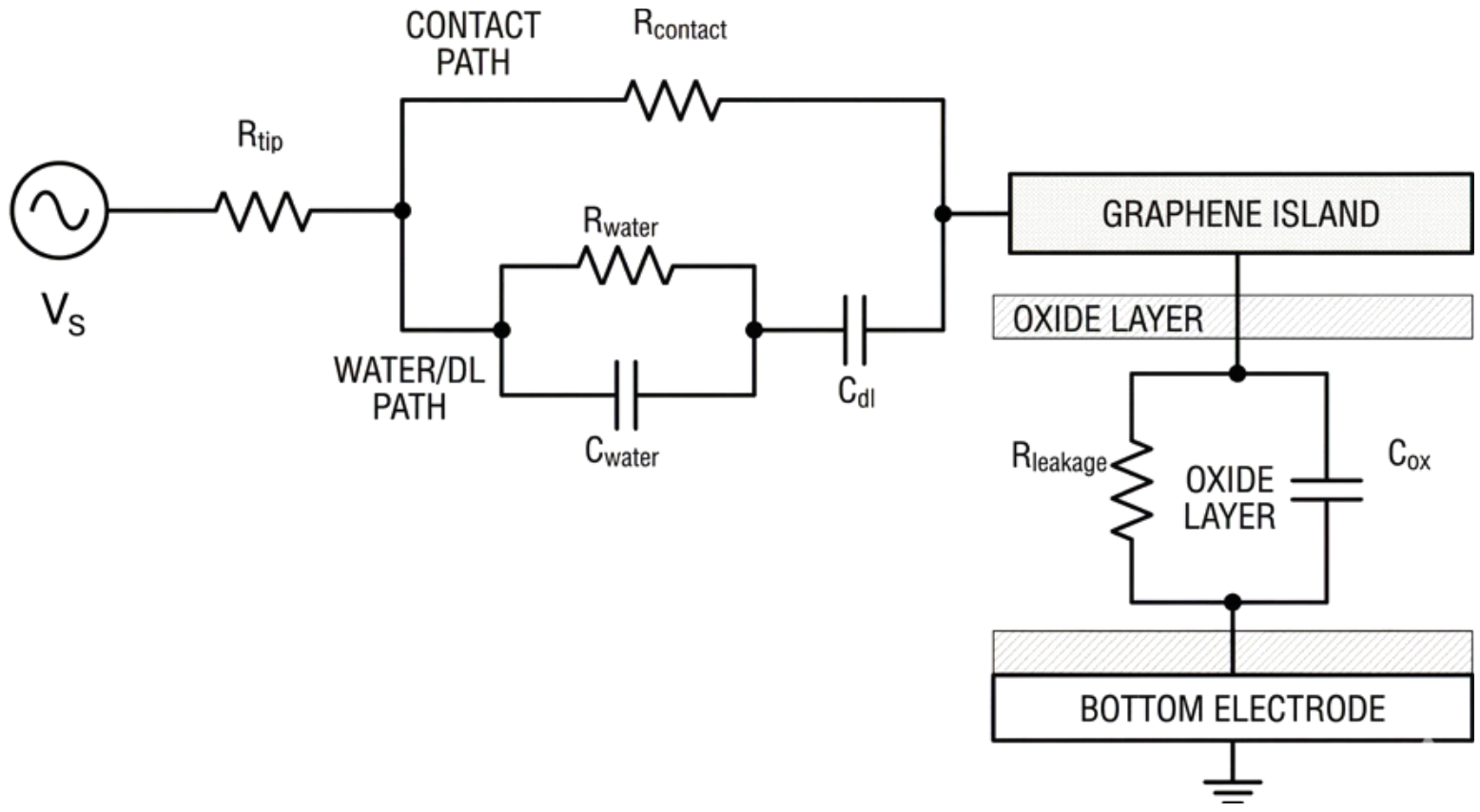

***Figure 9.*** *The equivalent circuit diagram illustrating the pathways via which current flows through the system. The current flowing through the water meniscus is what leads to etching of the graphene surface.*

Using this model, and with the assumption that the potential drop and therefore electric field between the voltage source and the graphene is what triggers the etching process, we can infer the following:

- Under DC or low-frequency operation, the pathways between the tip and graphene through the meniscus and between the graphene and the bottom electrode are open-circuit as a result of the capacitive elements, and the only current pathway which is open is the one at the contact point, which does not contribute to etching.
- As frequency is increased, the relative impedance of the pathway between the tip and graphene through the meniscus and between the graphene and the bottom electrode will vary. As an example, at 20 kHz, the impedance of the meniscus will be $R_{water}$ in parallel with $C_{water}$, which are 1 GΩ and around $10^4$ GΩ, respectively. This means that the dominant current through the water will be Faradic rather than displacement current. Ultimately, it is the Faradic current, associated with the physical movement of ions, which gives rise to the etching we observe.
- As the lateral size of the graphene is reduced, its impedance will increase and the voltage between the graphene and the bottom electrode will therefore increase, resulting in a decrease in the voltage between the voltage source and the graphene and a decrease in the electric field in the meniscus. In other words, as we reduce the size of a graphene island, it will become increasingly difficult to etch it, as seen in Figure 8.

Going beyond this circuit representation which describes the system as a whole and offers useful insight, we have developed a model to describe (i) the adsorption of water on all surfaces, (ii) the emergence and shape of a meniscus and then (iii) the distribution of electric field and potential in the region around the tip apex. This will allow us to understand the effect of RH, tip geometry, graphene island size and the reasons for etching only working for the case of an AC voltage and the graphene floating. Since water serves as the reaction medium in which field-driven oxidation of graphene occurs, an accurate description of its distribution around the contact point is essential. A detailed discussion is provided in the Supplementary Information.

In Figure 10, we show a set of results for the system, for a RH of 60%, which is the RH where the majority of the experiments were carried out. The applied voltage is 5.5 V at 20 kHz. In all cases where we plot the electric field, we are doing so at a height of 0.3 nm above the surface, so we can be certain that it is below the top of the meniscus for all values of RH we have considered. In the 2D XZ plot in Figure 10(a), the outline of the meniscus can be clearly seen, and the peak electric field is inside it. In the 2D XY plots of the electric field, there is a region approx. 8 nm across centred at the contact point within which the electric field is zero – this is inside the tip. We can see that the electric field is significantly enhanced immediately around the edge of the contact point, reaching values of over 4 V/nm at the peak, in a ring approx. 2 nm thick. This is not sufficient to cause ionisation of pure water, highlighting the importance of impurities in the adsorbed water layer. In the XY plot, we also indicate the extent of the meniscus at the height of 0.3nm as a dashed line, so we can see that the peak electric field is indeed occurring inside water, enabling electrolysis as long as the overpotential and electric field are sufficiently high.

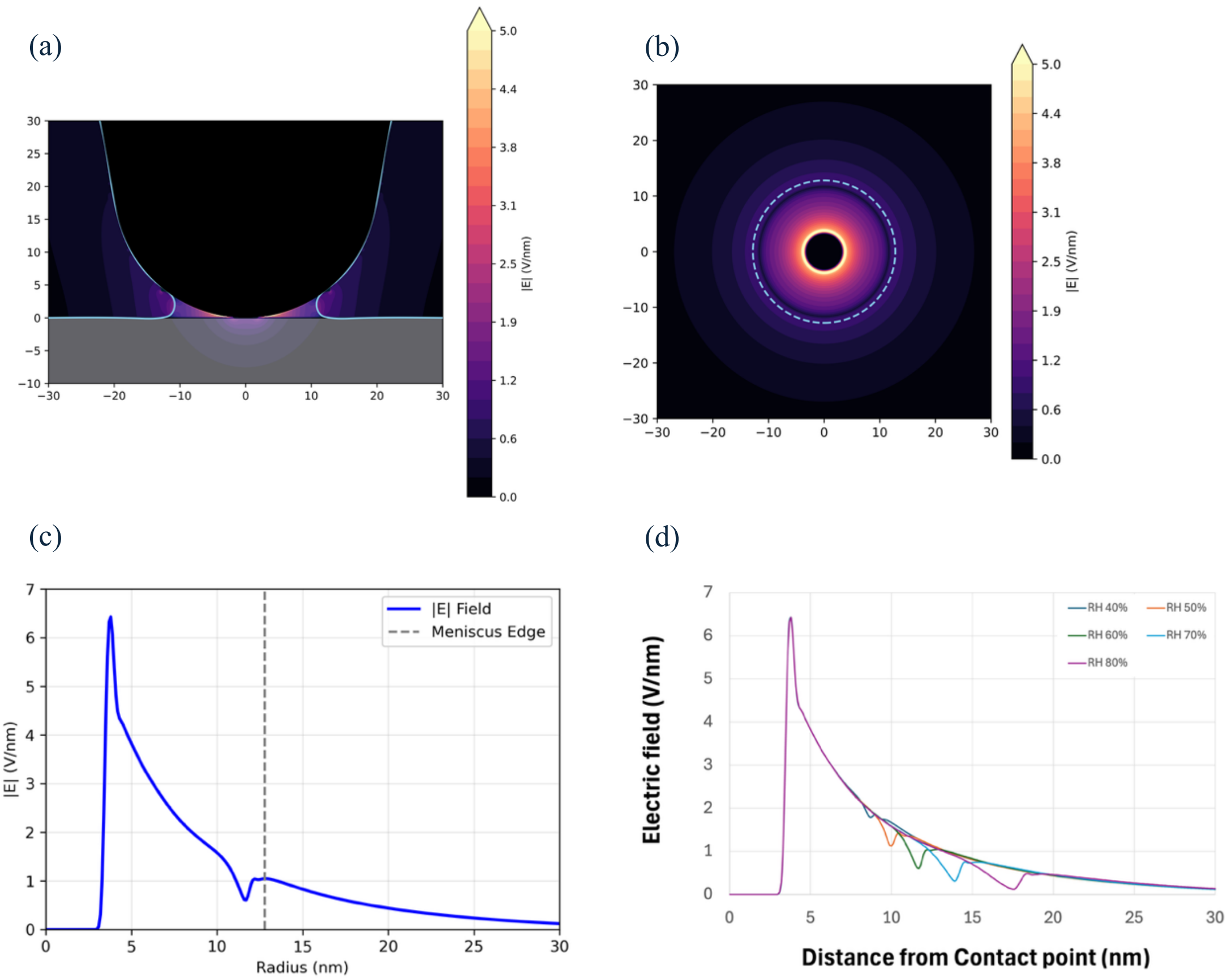

***Figure 10.*** *Plots of the electric field in the vicinity of the tip/graphene contact point for a RH of 60%, with the calculated meniscus shape (a) through the XZ plane, (b) in the XY plane at a height of 0.3nm from the top of the graphene surface, where the meniscus radius is indicated with a dashed line and (c) the electric field as a function of position, also at a height of 0.3nm, showing a peak value of 4.5 V/nm and a kink at the point where the meniscus ends, (d) plot of electric field as a function of position for RH varying in the range 40% to 80%.*

As RH is increased, we see that the peak electric field remains constant but the edge of the meniscus spreads out laterally, as shown in the plot in Figure 11(d). In all cases, there is a drop in field just inside the edge of the meniscus due to charging at the water-air interface.

The electric field also varies with the position of the tip relative to the edge of graphene. To explore this, we simulated a domain which extends from $x = -400$ nm to $+400$ nm. The domain contains a flake of graphene which extends all the way from the right edge of the domain at $x = +400$ nm to $x = -250$ nm, and from $x = -250$ nm to $-400$ nm there is just the underlying oxide. We have programmed the trajectory of the tip as it scans over the surface so that it is always just in contact with the top of the underlying surface, i.e. it moves upwards by an amount of 0.34 nm on a curved path as it moves over the edge. The field is somewhat weaker over $SiO_2$ than over graphene (by around 95%) and as we move the tip towards and then over the graphene edge, starting on the oxide, as shown in Figure 12, we see that as soon as the meniscus reaches the edge of the graphene (at $x = -275$nm), there is a small kink upwards in the peak field. More noticeable though is that there is a local enhancement of the field of around 40% when the tip is a few nm away from the edge, showing that the peak field is larger when starting on oxide and scanning over an edge than it is when starting on graphene. In Figure 11, we show the field distribution for three different tip positions- (a) far from the edge and on the oxide, (b) near the edge, at the position where the field is starting to increase and (c) on graphene and away from the edge, where the field is larger than on the oxide. In Figure 11(d), we show the peak field within the meniscus as a function of tip position where the enhancement is immediately apparent. Another point of note is that when etching is taking place, there is essentially an etched graphene edge all around the tip, and as it scans over the surface, creating a pattern, the field will be enhanced more than it was at the point when the voltage was initially turned on. In other words, as soon as etching starts and a hole is created, the field is significantly enhanced thereafter.

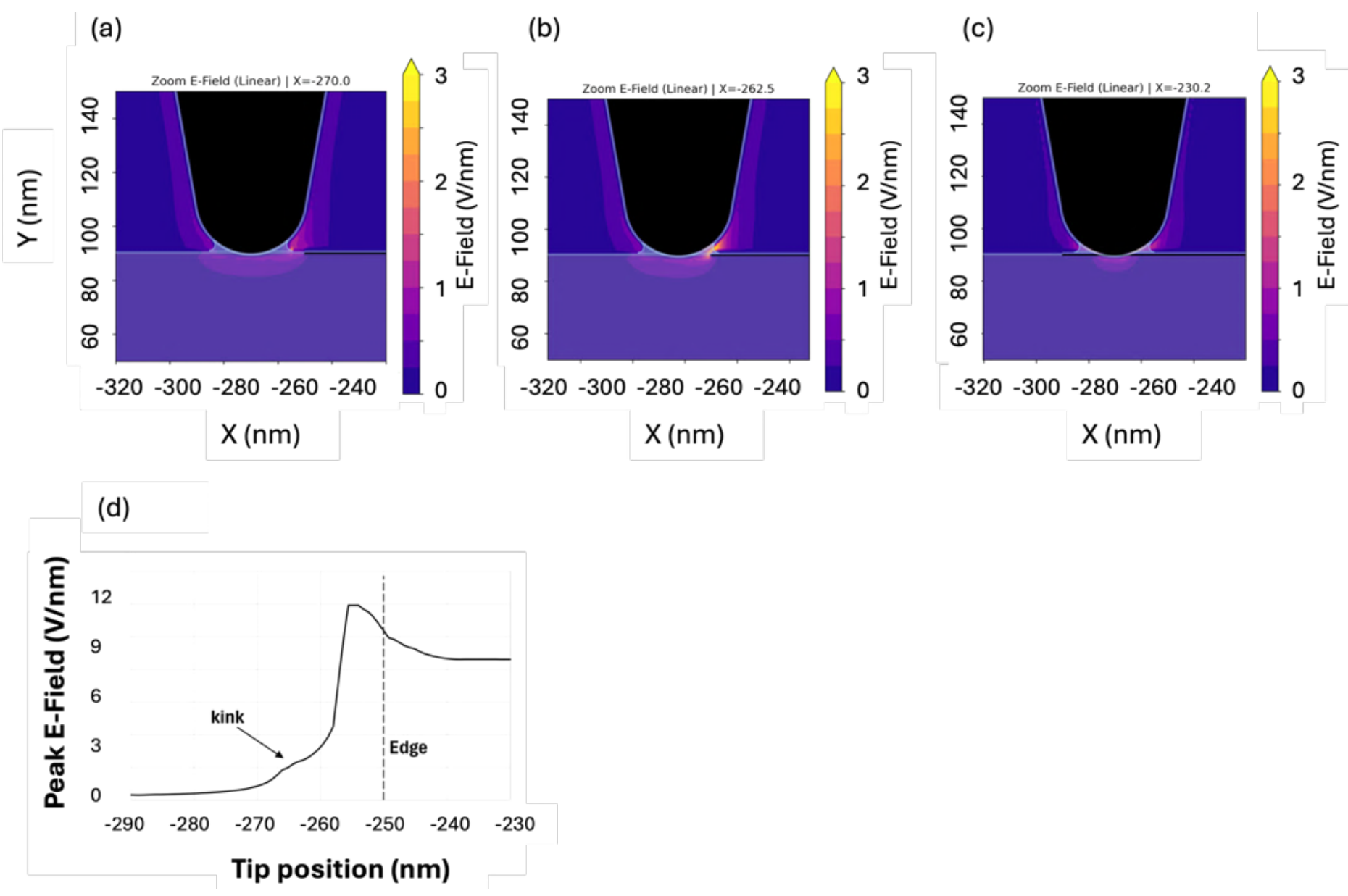

***Figure 11.*** *(a)-(c) 2D XZ plot of the Electric field for three different positions of the tip; (d) Peak electric field within the meniscus as a function of tip position. The vertical dashed line indicates the position of the graphene edge and the kink refers to the point at which the meniscus reaches the edge. The local field enhancement when the tip is around 5nm away form the edge is clearly visible.*

Our experiments have shown that as we reduce the size of an isolated graphene island, the etching process becomes less reliable, requiring a higher voltage. This is primarily why the etching did not work initially inside the island shown in Figure 8, until the tip reached the edge, where the aforementioned field enhancement took over. From the circuit model, as the capacitance of an island (denoted by $C_{oxide}$) reduces, the voltage drop across the meniscus will drop (as the island voltage will float up closer to the value of the tip voltage), hence the field in the water will drop. This island capacitance will scale as the island area, so we would expect as the side-length or radius of an island decreases, the voltage (and corresponding field) will drop. This is shown in Figure 12 where we plot the peak voltage and electric field as a function of island radius, where we see that even for an island 850 nm across (similar to that shown in Figure 8), the peak field has dropped by a factor of over 6 as compared to the case of continuous graphene. In this case, we have simulated a circular island to take advantage of the axisymmetric symmetry to enable faster computation.

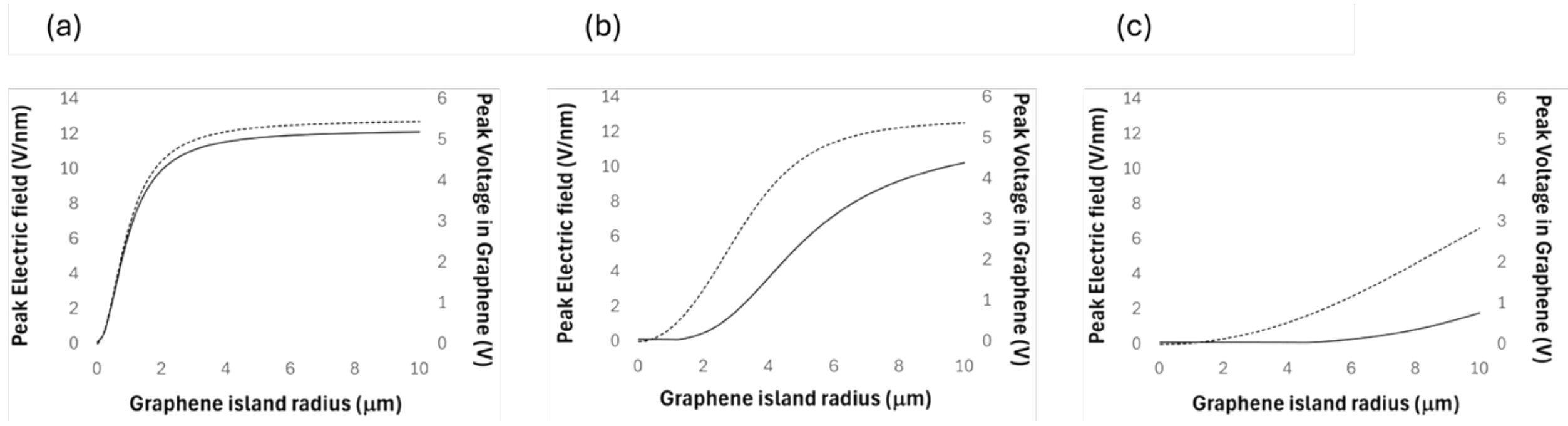

***Figure 12.*** *Voltage (dotted line) across and peak Electric field (solid line) within the meniscus as a function of graphene island size. RH = 60% and tip voltage is 5.5V at (a) 20 kHz, (b) 200 Hz and (c) 200Hz but with the water conductivity increased by a factor of 10 to 1 mS/m.*

The specifics as to the peak field vs frequency is highly dependent on the properties of the adsorbed water. As the thickness of these water layers increases, their conductance will also increase. but it will be highly dependent on the presence of impurities. Increasing conductivity will have the effect of suppressing the electric field, especially at lower frequencies and it will shift the balance of how much current flows through the resistance of the water (Faradic current) and through the capacitance (displacement current). The displacement current will be dependent on the capacitance of the island so will depend on size, and the above models (both the circuit model and the simulation) predict that increasingly conductive water/electrolyte will cause the field to be smaller for (a) lower frequencies and (b) smaller islands. This is evident in Figure 12 in terms of the attenuation of the field at low frequencies.

Using this etching technique, we then demonstrated how to fabricate a GNR structure. The starting point was a graphene microribbon connected between a pair of metal (Au) electrodes, followed by using the above technique to trim away the edges in a controlled way. As shown in Figure 13 (a), the tip was used to remove graphene in such a way as to leave a narrow ribbon connected to wider regions, which act as Source (S) and Drain (D), as indicated by the dashed lines. After the etching process, the topography image of one GNR of width 250nm is shown in Figure 13 (b). Confirmation of electrical isolation of the ribbon form the areas above and below was confirmed by C-AFM as before, as shown in Figure 13 (c) which shows that current can flow through the Source, Drain and GNR. The result of a CNP measurement, shown in Figure 13 (d) is unsurprising in that there is no band gap given the width. It also has a positive CNP, as is usually found from these wider GNRs.

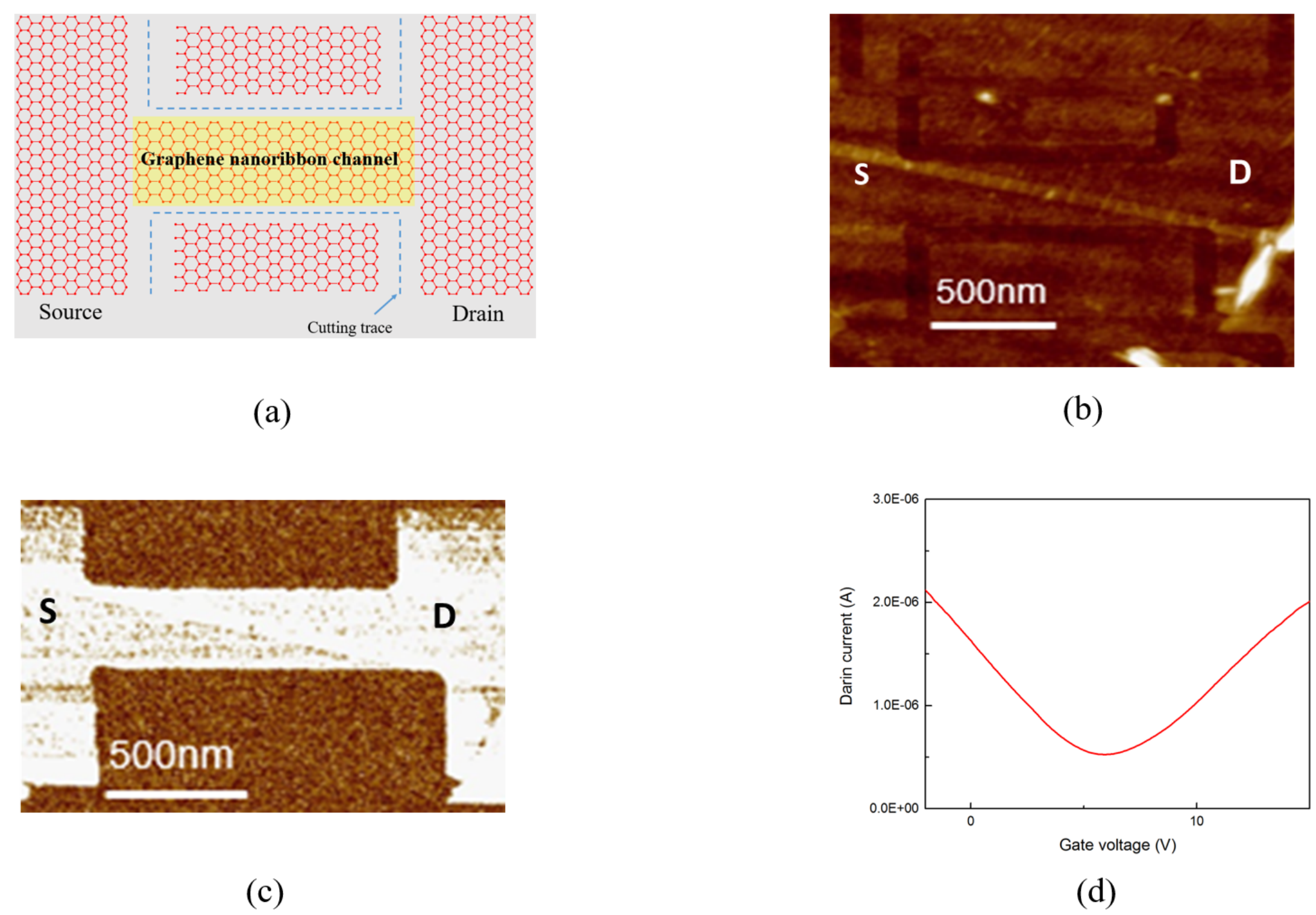

***Figure 13.*** *Illustration of the device fabrication and testing. (a) Schematic of the FET device fabrication. (b) Topography image of the process of cutting out the graphene channel. This image was scanned by AFM in tapping mode. (c) Current mapping image of the process of cutting out the graphene channel. This image was scanned by AFM in C-AFM mode. (d) The result of the CNP measurement.*

We then created a significantly narrower GNR device in the same way, but in this case with a width of less than10nm. The topography of this device as imaged using Scanning Electron Microscopy (SEM) is shown in Figure 14 (a). The result of a CNP measurement is shown in Figure 14 (b). We note two key points: (i) the CNP has shifted to a negative value in agreement with previous studies, and (ii) the distortion of the curve from a parabolic to a flattened curve shows that a bandgap has opened up, as expected from such a narrow GNR.

(a) (b)

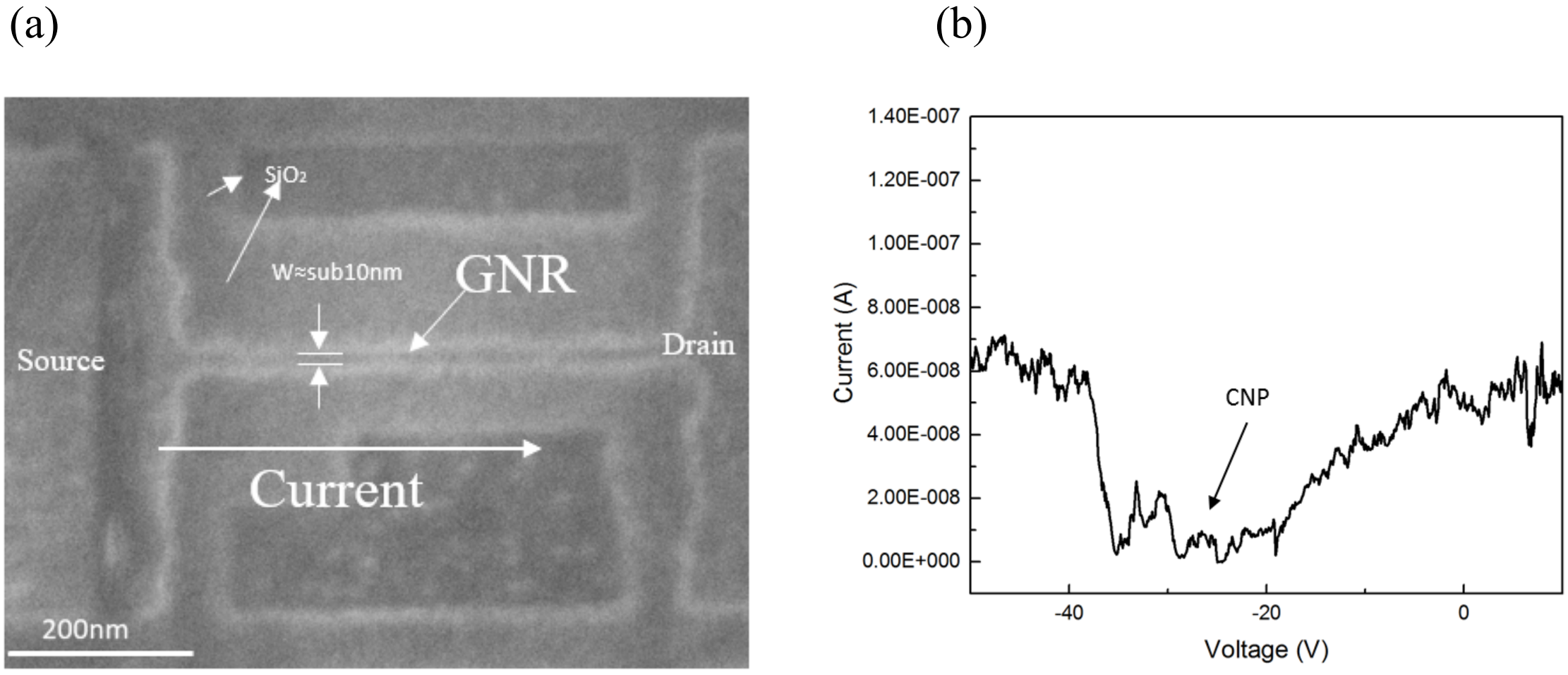

***Figure 14.*** *Illustration of the device fabrication and testing. (a) the topography of the sub-10nm device as shown by SEM. (b) CNP result of the device showing a bandgap and a shift towards a negative CNP.*

## Conclusion

In this article, we have presented our findings on the local AC electrochemical etching of graphene in an Atomic Force Microscope, with a proposed mechanism and model which can be used to explain how the process works experimentally. Our model can account for effects due to relative humidity, wettability of the graphene and AFM tip, conductivity of water and graphene, frequency and also the physical size of the graphene, as well as the geometry of the tip itself. Our results confirm that the tip remains in direct contact with the sample surface during patterning, rather than leaving a water-filled gap as had been proposed by others. Furthermore, we demonstrate that this method is stable and reliable for fabricating GNR-based devices with dimensions below 10 nm. Finally, we demonstrate the opening of a bandgap in a sub-10 nm GNR-based FET. The patterning approach demonstrated here can be extended to other 2D nanomaterials and their heterostructures, enabling the fabrication of geometrically well-defined samples for both fundamental research and next-generation device prototyping.

## Experimental

Monolayer graphene was grown on Cu foils (predominantly (110)-oriented) by chemical vapour deposition (CVD) and subsequently transferred onto p-type Si substrates coated with 90 nm using a wet-transfer process. In this study, CVD graphene field-effect transistor devices with pre-patterned metal contacts were supplied by Graphenea. Prior to lithographic processing, samples were cleaned sequentially in acetone and isopropyl alcohol.

AC-LAO was performed in contact mode using either an NT-MDT Solver Pro-M or a Park Systems XE-100 atomic force microscope. Relative humidity was controlled within a sealed enclosure housing both the AFM head and sample stage by introducing small water-filled containers, and was continuously monitored using a calibrated humidity sensor. Conductive probes (Multi75E-G; nominal tip radius <25 nm; half-cone angle, 10°) were used throughout. An AC bias was applied directly to the probe tip, delivered either through the microscope control electronics (NT-MDT) or externally via a Keysight EDU33212A waveform generator connected to the probe (Park Systems). Unless otherwise stated, the graphene channel was left electrically floating during patterning.

Current mapping was performed on the same instrument in conductive AFM mode. The system provides a maximum measurable current of 50 nA with a noise floor of ~1 pA. Simultaneous acquisition of topography and current signal in contact mode enabled spatial mapping of local conductance across the patterned regions.

CNP measurements were carried out in a Lake Shore probe station using a Keithley Instruments 4200 source measurement unit (SMU). Measurements were conducted under low vacuum conditions (~ $10^{-4}$ mbar) to minimise environmental adsorbates.

**Contributions**

C.D. supervised the project, performed the theoretical modelling and some of the experimental work, and revised the manuscript. Xiao Liu conceived the idea and concept of the paper, contributed to the theoretical analysis and interpretation, carried out most of the experiments, and wrote the manuscript.

# Supplementary Information

## Direct-write electrochemical nanofabrication of ultrasmall graphene devices

X. Liu[1] and C. Durkan[1,2]

[1]: Nanoscience, Department of Engineering, 11 JJ Thomson Avenue, Cambridge CB3 0FF, UK

[2]:corresponding author, email cd229@cam.ac.uk

**Supplementary Note 1| Physical assumptions**

**The water meniscus geometry and adsorbed water thickness**

We have used the Brunauer-Emmett-Teller (BET) model (*38*) which relates the volume, $v$ of adsorbed gas on a surface to the volume, $v_m$ of a single monolayer via the RH and the binding energy of the first monolayer as:

$$\frac{v}{v_m} = c\frac{RH}{(1-RH)[1+(c-1)RH]} \quad (1)$$

Where $c$ is a dimensionless constant, known as the BET (or affinity) constant related to the binding energy of that first monolayer of water on the graphene surface. Typical values of this constant vary significantly depending on the quality of the graphene, but range from around 1 to 200 for pristine graphene and graphene oxide, respectively (*39*). As we are dealing with CVD-grown graphene, it will be closer to pristine graphene, so we would expect the value for $c$ to be at the lower end (It will however be significantly larger on the $SiO_2$). The consequence of this is that the graphene is hydrophobic, so there will not be a continuous layer of water over the surface, and it will likely just form clusters. However, the high curvature at the tip–graphene contact promotes capillary condensation. The resulting curved liquid meniscus generates a Laplace pressure, which lowers the local vapor pressure and drives spontaneous condensation at sufficiently high relative humidity, leading to the formation of a nanoscale water build-up on the surface around the contact point. The consequence of the low affinity of water for graphene simply means that this meniscus will be tightly confined around the tip and does not spread out to wet the surface. This phenomenon of capillary condensation and the shape of the meniscus is captured by the Kelvin equation which describes the critical radius of curvature of the meniscus, $r_k$, which is the outer boundary of the water layer that forms around the tip and at the surface at the contact point:

$$r_k = c - \frac{2\gamma V_m}{RTln(RH)} \quad (2)$$

where $\gamma$ is the surface tension of water, $V_m$ is the molar volume of water, $R$ is the gas constant and $T$ is temperature(*40*).

We have combined these physics elements along with a finite-difference solver on a uniform grid to determine the electric field and potential using a successive over-relaxion method. We solve the Complex Poisson equation by solving the continuity equation for current in a lossy dielectric. We map the complex permittivity of each material type, i.e. air, water, graphene, oxide and metal to its specific location, and employ anywhere between 1000 and 50,000 iterations to find a stable solution, depending on the specific location of the tip. More iterations are required when the tip is on graphene, due to the high fields and high field gradients there. A schematic of the system being modelled is shown in Figure S1.

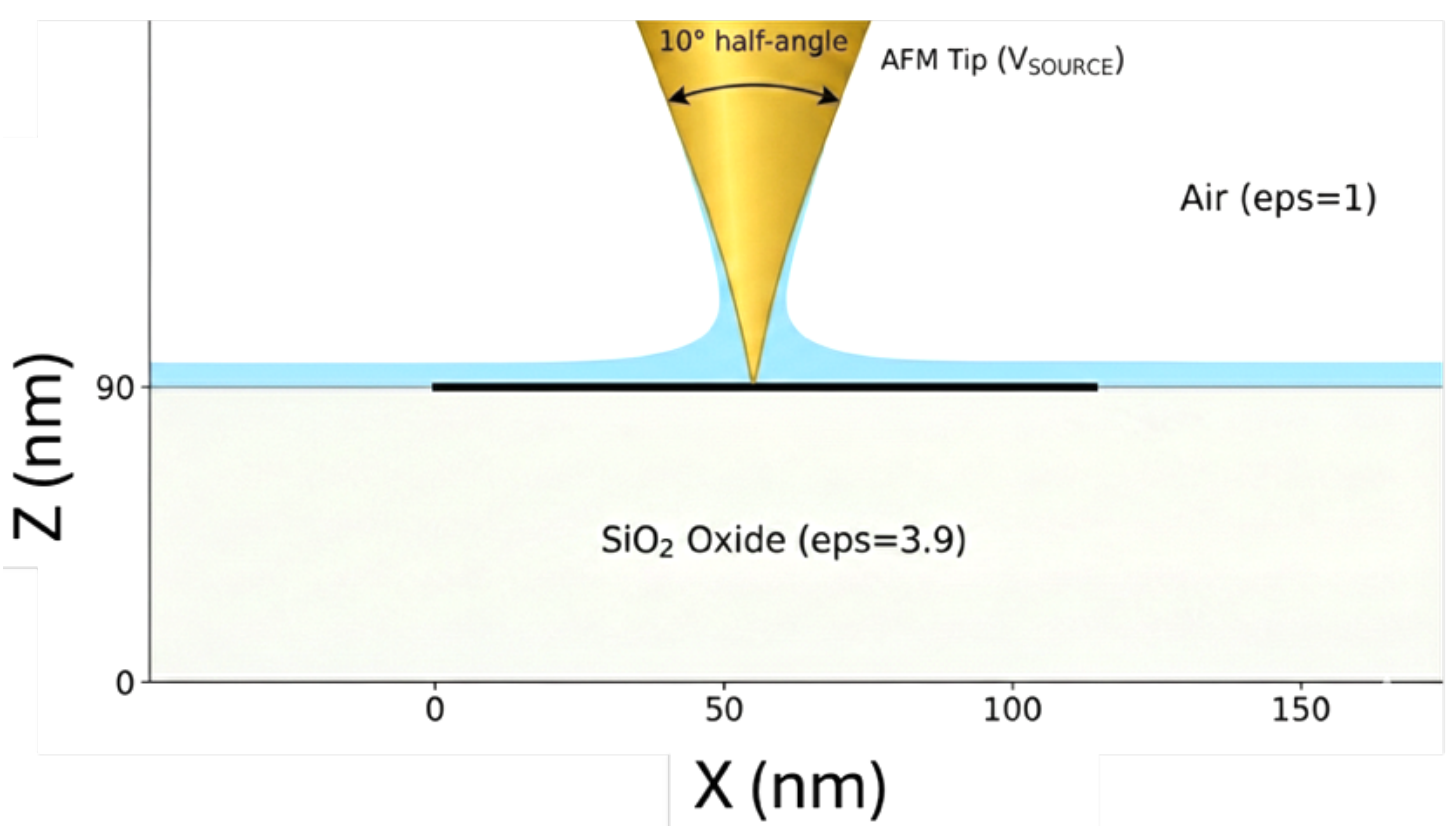

*Figure S1. The system being modelled.*

In Table S1, we list the parameters which are built into the model, all of which can be altered to change the geometry or materials. We are implicitly assuming that the complex relative permittivity of water is given by $\varepsilon_{water} = \varepsilon_r - j\frac{\sigma}{\omega\varepsilon_0}$.

| Parameter | Value |
|---|---|
| BET constant | 100 |
| Monolayer thickness | 0.3 |
| Surface tension of water | 72 mN/m |
| Conductivity of water, $\sigma$ | $10^{-3}$ to $10^{-4}$ S/m |
| DC relative permittivity of water, $\varepsilon_r$ | 80 |
| Tip radius | 10 nm |
| Tip cone half angle | 10 degrees |
| Oxide thickness | 90 nm |

***Table S1**. Parameters used in finite-difference model.*

This is a full 3D model which is primarily set in a Cartesian co-ordinate system, when looking at square or rectangular islands (such as the one created experimentally in Figure 8) or when exploring the effect of tip position relative to an edge. For cases where we are looking at the shape of the meniscus or not moving the tip, we can revert to an axisymmetric model which is essentially 2D, with a reduction in computing cost of approx. 99%. Given that the Hertzian contact radius of an AFM tip with a sample under these conditions will be of order 1nm [Based on Hertzian contact mechanics, the contact radius scales as $a \sim \left(\frac{FR}{E^*}\right)^{1/3}$, so for typical AFM parameters (*41*)(*42*) the grid needs to be small enough to capture the extreme gradients in potential and field. The grid size is therefore set to be in the range 0.1nm to 1 nm, depending on the specific region of interest we are focusing on. In any case, we typically have of order $10^9$ elements.

**Supplementary Note 2| Electrochemical reactions during AC bias cycling**

We have experimentally determined that the typical contact resistance between the tip and CVD graphene is around 2 MΩ. Additionally, the sheet resistance of CVD graphene is typically in the range $350 - 800\ \Omega \text{sq}^{-1}$ (*43*). Although the RH is controlled by placing small containers filled with deionised water near the sample, contaminants in the air and already on all surfaces mean that the adsorbed water on the graphene surface (and on the tip surface also) is anticipated to have a finite conductivity as mentioned earlier. Whether the electric field is large enough to induce splitting of the water is something we will explore later. In line with what would be expected for Local Anodic Oxidation (LAO) (*13*, *14*, *35*), the effect of the electric field will be to drive ions within the meniscus towards the graphene surface and the tip, and as long as the overpotential on the graphene is sufficient (typically needs to be at least 1.229V(*44*)), etching will proceed as follows over the course of the AC voltage cycle:

1. **Tip negative, graphene positive**: In this case, any negative ions within the water, including $OH^-$ and $O^{2-}$ will be driven towards the graphene surface, enabling the following reaction to take place there:

   $$C_{Solid} + 2H_2O \rightarrow CO_2 + 4H^+ + 4e^-$$

   In other words, the solid graphene is transformed into $CO_2$ gas. This reaction will be confined within the nanometric volume occupied by the meniscus. The 4 electrons thus released then go into charging up the graphene, and will be subsequently removed on the positive half of the voltage cycle. Under DC conditions, this removal of electrons does not happen, so the island will charge up, and once it reaches the tip potential, the current will cease, and the etching reaction will self-terminate. The timescale in which this happens will be determined by the resistance of the contact and the graphene, as well as the capacitance of the double layer and the graphene/oxide/back electrode system, which we can express via an *RC* time constant.

The voltage provides the thermodynamic energy to the system to enable the reaction to proceed, and the rate of that reaction will be determined by the field strength, as it delivers the current to/from the surface. The graphene, being positive is therefore the anode in the reaction, and the reaction above which is at the root of the etching process is known as local anodic oxidation, or LAO (*35*)(*40*).

Simultaneously, at the tip, the following reaction takes place:

$$2H_2O + 2e^- \rightarrow H_2 + 2OH^-$$

The $OH^-$ ions thus generated add to any pre-existing ions arising due to impurities in the water and participate in the etching reaction.

2. **Tip positive, graphene negative**: In this case, any positive ions, (primarily $H^+$) will be driven towards the graphene. Under these conditions, the Carbon will no longer be oxidised, but instead water will be reduced on the graphene surface(*45*):

$$2H_2O + 2e^- \rightarrow H_2 + 2OH^-$$

Whereas on the tip, there will be a corresponding oxidation reaction:

$$2H_2O \rightarrow O_2 + 4H^+ + 4e^-$$

This will create an oxygen-rich environment around the tip which may lead to the tip becoming oxidised, depending on the material it is made from.

In both halves of the cycle therefore, $OH^-$ ions are being produced which are then accelerated towards the graphene during the part of the cycle where the tip voltage is negative.